\documentclass[acmtog]{acmart}

\AtBeginDocument{%
  \providecommand\BibTeX{{%
    \normalfont B\kern-0.5em{\scshape i\kern-0.25em b}\kern-0.8em\TeX}}}

\usepackage{multirow}
\usepackage{graphicx}
\begin{document}

\title{Security and Privacy Problems in Voice Assistant Applications: A Survey}

\author{Jingjin Li}
\email{ginger.li@my.jcu.edu.au}
\orcid{0000-0003-4390-9479}
\authornotemark[1]
\affiliation{%
  \institution{James Cook University}
  \streetaddress{1 James Cook Dr, Douglas QLD 4811}
  \city{Townsville}
  \state{QLD}
  \country{AU}
  \postcode{4810-4811}
}

\author{Chao chen}
\affiliation{%
  \institution{RMIT University}
  \streetaddress{124 La Trobe St, Melbourne VIC 3000}
  \city{Melbourne}
  \country{AU}}
\email{chao.chen@rmit.edu.au}

\author{Lei Pan}
\affiliation{%
  \institution{Deakin University}
  \city{Geelong}
  \country{AU}}
 \email{l.pan@deakin.edu.au}

\author{Mostafa Rahimi Azghadi}
\affiliation{%
 \institution{James Cook University}
 \streetaddress{1 James Cook Dr, Douglas QLD 4811}
 \city{Townsville}
 \state{QLD}
 \country{AU}}
 \email{mostafa.rahimiazghadi@jcu.edu.au}

\author{Hossein Ghodosi}
\affiliation{%
  \institution{James Cook University}
  \streetaddress{1 James Cook Dr, Douglas QLD 4811}
  \city{Townsville}
  \state{QLD}
  \country{AU}}
\email{hossein.ghodosi@jcu.edu.au}

\author{Jun Zhang}
\affiliation{%
  \institution{Swinburne University of Technology}
  \streetaddress{John St, Hawthorn VIC 3122}
  \city{Melbourne}
  \state{VIC}
  \country{AU}}
\email{junzhang@swin.edu.au}

\renewcommand{\shortauthors}{Li, et al.}

\newcommand{\myauthornote}[3]{{\color{#2} [{\sc #1}: #3]}}
\newcommand\sky[1]{\myauthornote{Sky}{red}{#1}}

\begin{abstract}
Voice assistant applications have become omniscient nowadays. Two models that provide the two most important functions for real-life applications (i.e., Google Home, Amazon Alexa, Siri, etc.) are Automatic Speech Recognition (ASR) models and Speaker Identification (SI) models. According to recent studies,  security and privacy threats have also emerged with the rapid development of the Internet of Things (IoT). The security issues researched include attack techniques toward machine learning models and other hardware components widely used in voice assistant applications. The privacy issues include technical-wise information stealing and policy-wise privacy breaches. The voice assistant application takes a steadily growing market share every year, but their privacy and security issues never stopped causing huge economic losses and endangering users' personal sensitive information. Thus, it is important to have a comprehensive survey to outline the categorization of the current research regarding the security and privacy problems of voice assistant applications. This paper concludes and assesses five kinds of security attacks and three types of privacy threats in the papers published in the top-tier conferences of cyber security and voice domain.
\end{abstract}





\maketitle

\section{Introduction}

\begin{figure*}[!ht]
\centering
  \includegraphics[width=\textwidth]{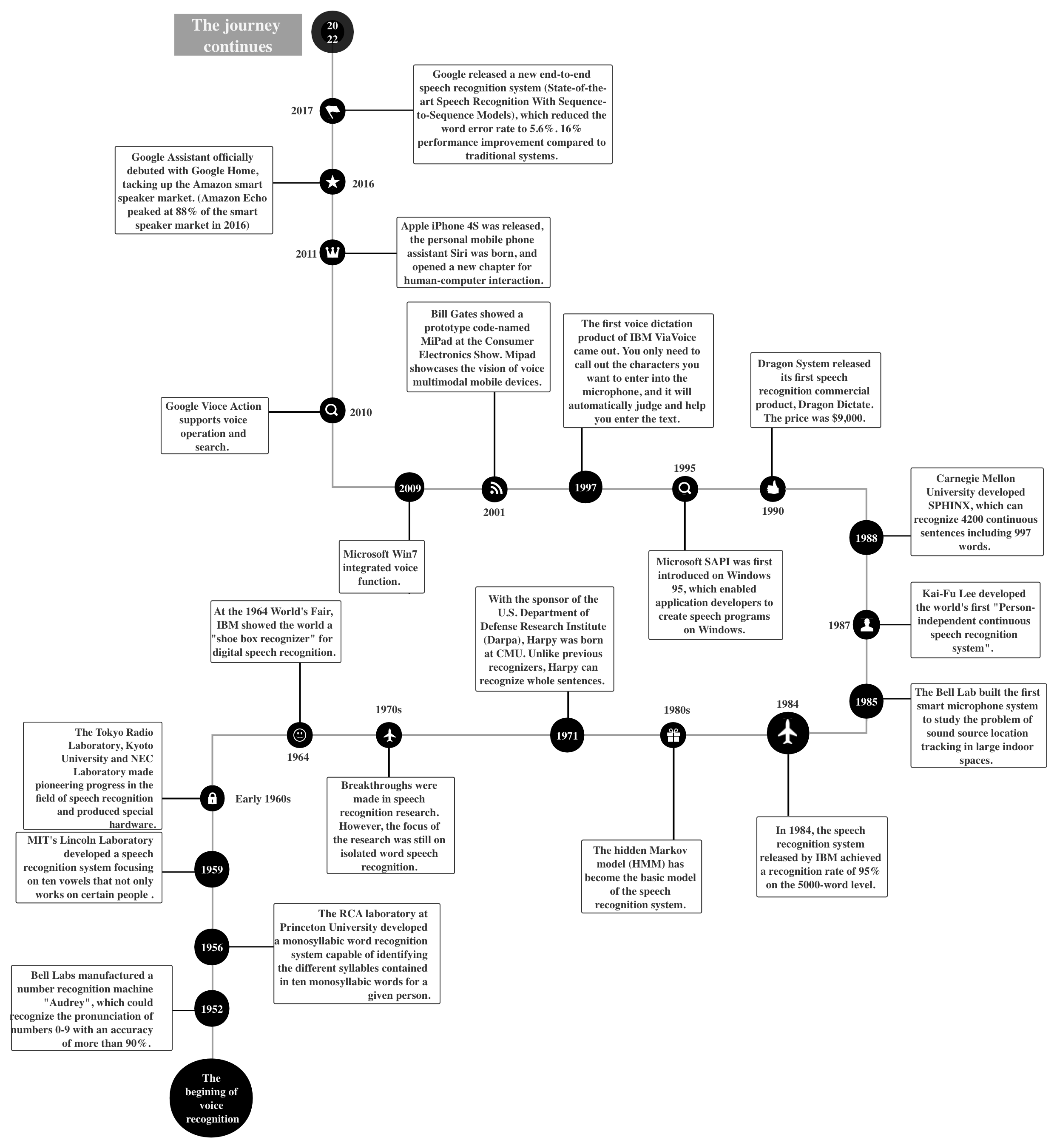}\\
  \caption{The timeline that outlined the development and progress of voice assistant applications.}
  \label{fig:timeline}
\end{figure*}

Naturally, people communicate through voice. Three technologies have been proposed to facilitate human-computer interaction through voice, including automated speech recognition (ASR), natural language processing (NLP), and speech synthesis (SS). NLP enables machines to comprehend human intents, and SS enables machines to talk. The research of voice assistant applications started in the 1950s with continuous refinement. Figure \ref{fig:timeline} shows the progress of voice assistant applications over time. The Hidden Markov Model (HMM) method, a statistical model-based approach, has increasingly taken the lead in voice recognition research since the 1980s. As voice recognition technology thrives with the more advanced deep learning algorithm, it integrates with more and more devices. In 2011 when the iPhone 4S was released, the world's first mobile phone personal voice assistant Siri was known and opened a new chapter for voice assistant applications.

Nowadays, voice assistant applications prosper with a large portion of the market. The voice assistant has brought huge convenience to our daily life and greatly changed how humans interact with computers. Almost every smart device has a built-in voice assistant. Because of the development and wildly use of voice assistant applications, privacy and security problems emerge. Often, users may not even notice their conversation has been recorded or mistakenly awoken by the voice assistant. Taking Siri as an example, several times it heard other people saying ``Are you serious?" or ``a series of ...", it just started to voice recognition and write down what it heard on my screen. A voice assistant is easily activated by accident, which malicious attackers could exploit. Plausible but severe threats include bank transfers, buying virtual products, fabricating messages to your close friends or families asking for money, stealing your credential information, and many others. Large financial and emotional losses may occur if voice assistant applications are breached. Thus, the security and privacy problems within voice assistant applications should be followed with interest. Users become more focused on taking complete control of their voice assistant applications, knowing the potential attack and defense methods.

With many papers published in recent years about new attack techniques and defense means, a comprehensive outline of their development is needed. However, there are two surveys in the voice assistant security and privacy domain. Cheng et al.~\cite{Survey1} focused on security and privacy problems that voice assistants have in using acoustic channels. Four main problems were included in Cheng et al.~\cite{Survey1}. They are access control loss, acoustic DoS attack, voice privacy loss, and malicious use of acoustic sensing. Compared to their survey, this survey also covers none acoustic channel attacks. Also, their survey only covers research before the end of 2020, and this survey covers more recent research published in top security conferences until 2022. In this survey, during 2020-2022, quite a few papers about new side-channel attacks were researched and had impressive results. Another difference is that this survey divided the attacks towards ASR and SI models. Because ASR and SI are the two major functions that voice assistant applications have, some applications have both, and some have only one. It is useful for users to know which function their applications have and what type of threats correspond to each application. The other survey by Yan et al.~\cite{Survey2} covers almost the same security and privacy problems mentioned in Cheng et al.~\cite{Survey1}. However, it also does not cover any research after 2020. Also, Yan et al.~\cite{Survey2} categorize defensive methods from a system designer's perspective, which differs from this survey and Cheng et al.~\cite{Survey1}. In this survey, defensive methods are introduced regarding specific attacks. Through this survey, some defensive methods are effective in multiple types of attacks, and some mitigate one type of attack but are prone to another type of attack, which helps the users or producers have more comprehensive information when choosing the defensive methods.

This survey aims to make a comprehensive and clear outline of the security and privacy issues of voice assistant applications. The papers included in this survey are from the top four cyber security conferences and Interspeech, a conference focusing on the speech domain. The aspects that are included are as follows:

\begin{enumerate}
  \item Technical attacks that were targeted towards ASR models and SI models. Including machine learning attacks that targeted the software, frequency modulation that exploits the hardware, malicious skills hidden in the third-party market and policy loopholes that were not refined quickly enough to catch up on the development of voice assistants.
  \item Defensive methods have been researched and proven effective in ASR and SI models. Usually, the defensive means can be divided into detection and prevention. Some methods may provide both means. 
  \item We have security and privacy issues when using voice assistant applications beyond technical threats. With more and more younger users, third-party regulations and policies should be refined.
\end{enumerate}

\noindent \textbf{Contributions.} This paper provides a comprehensive summary of technical attacks with impressive experiment results and feasible defensive methods corresponding to each attack. Nontechnical threats in the voice assistant application market are included to safeguard the user. The contributions are concluded as follows:

\begin{itemize}
 \item From a user's standpoint, this survey is, as far as we are aware, the most thorough investigation of voice assistant application security. Our study includes both market policy issues and technology risks. We provide a comprehensive overview of the state of the art, development, major difficulties, and future prospects for voice assistant application security research based on a thorough literature review of pertinent attacks and countermeasures.
 \item We classify pertinent assaults by attack techniques and structure the attack literature according to the voice assistant's systems. In order to properly identify, comprehend, and analyze the security risks against voice assistants, the organization assists in bridging the gap between a large category of seemingly unrelated attacks and vulnerabilities.
 \item To systematize the countermeasures against various attacks, we base them on defensive tactics. We present a qualitative evaluation of existing solutions by the installation cost if the defense requires additional devices, usability, and security and make useful recommendations in order to help users select protection based on the type of danger they may encounter.
\end{itemize}

The remaining portions of this essay are structured as follows: The introduction to voice assistant applications in Section \ref{sec:vaa} is brief. The taxonomy of assaults on ASR and SI models as well as the taxonomy of countermeasures that may be applied to ASR and SI models are also introduced in Section \ref{sec:vaa}. The attacks that take advantage of the voice assistant's ASR function's weaknesses are described in detail in Section \ref{sec:asr} along with the corresponding defenses. The attacks against SI models are described in detail in Section \ref{sec:si}, along with systematized defense tactics that can stop or at least slow them down. The security and privacy issues outside of technological assaults are summarised in Section \ref{sec:pri}. In Section \ref{sec:cha}, we go through issues with the current research and potential future approaches for voice assistant applications. The survey is concluded in Section \ref{sec:con}.

\section{Preliminaries of Voice Assistant Applications}
\label{sec:vaa}
This section provides background information on voice assistant apps, including a definition of key terms, a list of categories, and a description of the process for each type of voice assistant application. 

\subsection{Voice Assistant Components and Speech Recognition Workflow}

\begin{figure*}[!ht]
\centering
  \includegraphics[width=\textwidth]{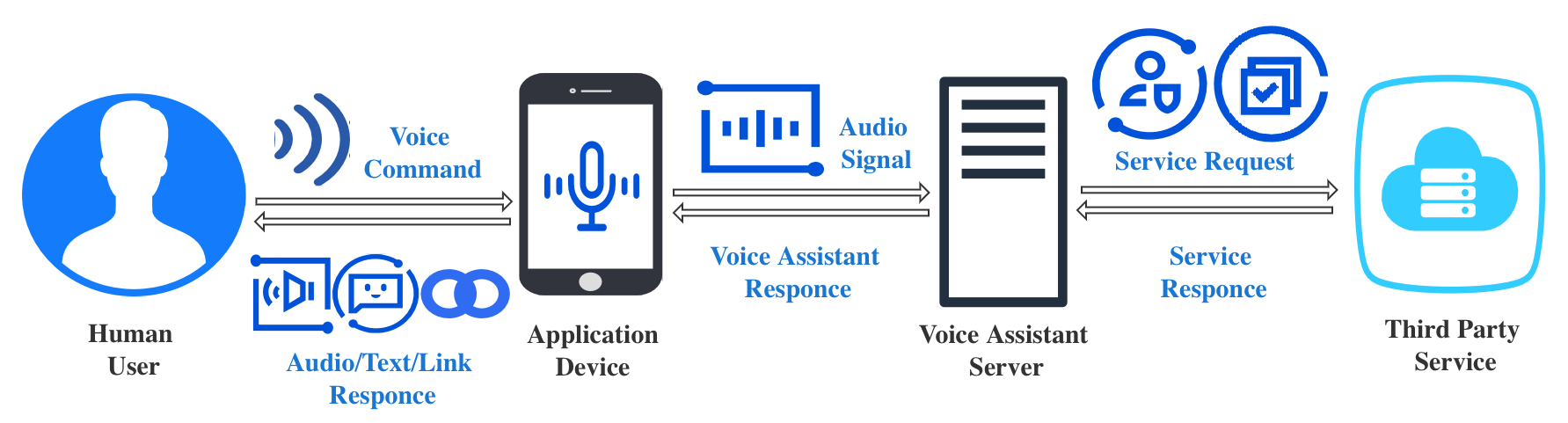}
  \caption{Workflow of voice assistant application service.}
  \label{fig:workflow}
\end{figure*}

There are two kinds of voice assistant models --- automatic speech recognition (ASR) and speaker identification (SI). As shown in Figure \ref{fig:workflow}, the first step in creating a voice recognition model is translating the spoken language into text. Speech recognition is much more challenging to solve than machine translation. A machine translation system's input is usually printed text that differentiates between individual words and word strings. The voice input used by a speech recognition system is far more complicated than written text and spoken language, especially with ambiguity. When two people communicate, they frequently infer the term in the conversation in the context and often read a lot of latent information from the tone, facial expressions, and gestures the other party uses. The speaker regularly rectifies what has been said and repeats important material by rephrasing. It is challenging to train an automated system to detect and comprehend speech. To provide a compact digital representation of the sound wave, each sampled value is quantized throughout the speech recognition process. A feature vector characterizing the spectral content is retrieved for each frame from which the sampled values are situated in overlapping frames. The words that the speech represents are identified based on the features of the voice signal. The five steps that make up the voice recognition process are described as follows: 

\noindent \textbf{Step 1. Voice Signal Acquisition}

Voice signal acquisition is the foundation of voice signal processing. A voice signal acquisition system typically receives inputs through a microphone. Subsequently, the sound wave is transformed from a voltage signal by the microphone to a digital signal handled by an A/D device like a sound card. Voice signal acquisition and processing systems based on single-chip microcomputers and DSP chips are utilized extensively for unfavorable on-site conditions, limited space, and numerous specific equipment. The essential hardware for voice assistant apps includes sound cards, speakers, microphones, and many alike. Sound cards are crucial to process voice signals through signal filtering, amplification, A/D conversion, and D/A conversion. Modern recording software tools activate the sound card to harvest voice signals as voice recordings.

\noindent \textbf{Step 2. Speech Signal Pre-processing}

After collecting the speech signal, pre-processing operations must be completed, including filtering, A/D conversion, pre-emphasis, and endpoint detection. Filtering primarily serves the two goals of preventing aliasing interference and suppressing the 50 Hz power frequency interference. The voice analog signal is converted into a digital signal via A/D conversion. The signal is quantized during A/D conversion, and the quantization error, also known as quantization noise, is the difference between the quantized signal value and the original signal value. Pre-emphasis processing aims to improve the signal's high-frequency content, flatten its spectrum, and maintain its full frequency range from low to high frequency. Endpoint detection involves extracting the beginning and conclusion of speech from a speech-containing signal. Effective endpoint identification removes background noise in silent periods. Two popular approaches work on different features --- time-domain features and frequency-domain features. The time domain feature approach uses the voice volume and zero-crossing rate to identify endpoints with the advantage of a minimal amount of calculation. However, the time domain feature approach often leads to incorrect evaluation of air sounds, and differing volume calculations will also result in varied detection outcomes. On the other hand, the frequency domain feature technique uses variations in the sound spectrum and entropy detection to identify speech at a high computation cost.

\noindent \textbf{Step 3. Feature Parameter Extraction of the Speech Signal}

The frequency of human speech is below 10 kHz. Shannon's sampling theorem requires the sampling frequency to be at least twice the maximum speech frequency present in the speech signal. The signal is often broken into blocks (also known as frames). Frames should overlap each other to prevent losing crucial information. Microphones collect waveforms of sound. It is important to extract distinctive information to separate the words from the collected data. Techniques for linear predictive coding are frequently employed to extract voice components. The fundamental tenet of linear predictive coding is that speech signal sampling points are correlated so that a linear combination of numerous previous sampling points helps predict the values of the present and subsequent sampling points. The linear prediction coefficient is calculated to reduce the mean square error between the anticipated and actual values. 

\noindent \textbf{Step 4. Vectorization}

Vector quantization (VQ) is a data compression and coding method. In scalar quantization, a dynamic range is split into several sub-intervals, where each sub-interval has a representation value. This representative value is used to determine the value for an input scalar signal that falls inside the sub-interval during quantization. Due to scalar quantization, the semaphore is a one-dimensional scalar. VQ transforms a scalar into a one-dimensional vector from the perspective of linear space to quantify the vector. VQ separates the vector space into numerous little sections. A representative vector replaces the vectors in the section during quantization for each small section. VQ integrates various scalar values into a vector (or feature vector generated from a frame of speech data) to provide overall quantization in multi-dimensional space and enable data compression with minimal information loss. In a hidden Markov model, the input observation symbol can alternatively be the vector quantized feature vector.

\begin{figure*}[!ht]
\centering
  \includegraphics[width=\textwidth]{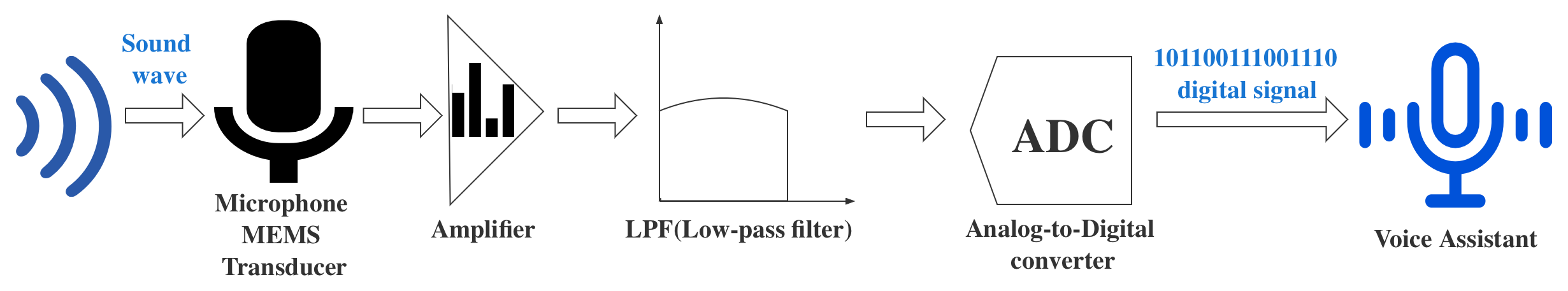}
  \caption{Microphone components and voice signal capturing and pre-processing workflow}
  \label{fig:preprocessing}
\end{figure*}

\noindent \textbf{Step 5. Speech Recognition}

A typical speech recognition task is recognizing words and phrases because words are sequences of letters. A recognition system receives feature parameters from the speech signal as the input, like the LPC predictive coding parameters. Using Bayesian decision-making with maximum likelihood, three typical approaches are used in speech recognition: template matching, stochastic model, and probabilistic parsing.
\begin{itemize}
 \item In template matching, a template is generated and stored while a user pronounces each phrase during the training stage. Each template in the template library is a feature vector. During the recognition stage, the input speech's feature vector sequence is iteratively compared to each template in the template library for the best match. 
 \item The hidden Markov model (HMM) is the most popular method among stochastic models. HMM is a time-varying process that transits from one reasonably stable feature to another characteristic. With adequate time, the speech signal's properties gradually stabilize. 
 \item Probabilistic parsing is used for continuous voice recognition across broad length ranges. While individuals speak the same phonetics, significant differences in the corresponding spectrograms and their modifications exist among individuals. 
\end{itemize}
Last but not least, several other voice recognition techniques exist especially artificial neural network-based approaches for voice recognition, including the BP neural network, the Kohcmen feature mapping neural network and other networks with deep learning. 

\subsection{Speaker Identification Workflow}

The Speaker Identification (SI) system is often referred to as speaker recognition. SI consists of two stages: speaker identification and speaker confirmation. Speaker identification is a one-to-one mapping, and speaker confirmation is a many-to-one mapping. SI determines whether multiple speakers are present in a record and validates a speaker's identity by analyzing and processing the speech signal of the speaker. SI creates a reference template or model by extracting unique characteristics from the original voice signal before recognizing a speaker according to the predetermined criteria. In a SI task, the system extracts the speaker's personality traits by averaging the semantic information in the speech signal, emphasizing the individual's distinctive characteristics; in a speech recognition task, the system normalizes the differences between different people's speeches as much as possible. The waveform of the speaker's speech reflects differences in pronunciation organs and habits, revealing each person's speech as a distinct personal trait that serves as an objective assurance of the speaker's identity.

Depending on the speaker numbers, speaker identification has two categories: closed set and open set. A closed set SI requires reflecting the number of speakers in the set to a closed set; on the contrary, an open set SI requires disregarding the number of speakers. Only a comparison and judgment between a reference model and the test speech are required for validation. 

Speaker identification may be broken down into three groups: text-related, text-independent, and text-prompted. The speaker's pronunciation of essential words and phrases is used as a training text by text-related SI, and the same information is uttered during recognition. The recognition object is a free speech signal, and the text-independent speaker identification technique does not define speech content during training or recognition. 




The training stage and the recognition stage are the two key phases. A template or model of each speaker is created during the training phase using feature extraction and the training corpus for each speaker in the speaker set. The speech to be recognized is broken down into its component characteristics at the recognition stage and compared to the template or model created during the system training. In speaker identification, the recognition outcome is the speaker corresponding to the model with the highest predicted speech similarity. Decide speaker confirmation by determining if the similarity between the test tone and the claimed speaker's model is higher than a predetermined threshold. The following fundamental issues affect the SI system's realization: 
\begin{enumerate}
  \item Preprocessing speech signals and feature extraction or extracting parameters can describe speaker characteristics. 
  \item Establishing the speaker model and establishing the model's training parameters. 
  \item Calculating the test speech's similarity to the speaker model. 
  \item Identification and technique for choosing. Confirmation or identification of the speaker.
\end{enumerate}

Three categories can be used to implement SI: 
\begin{enumerate}
 \item Template matching --- A reference template is a set of feature vectors to characterize the sequence of feature vectors. During the training process, feature vectors are extracted from the training sentences of each speaker to extract the feature vector sequence. During identification, a subject's template is compared with each reference template. The outcome of matching is frequently the accumulated distance between the feature vectors, measured as part of the matching process. VQ and dynamic time normalization (DTW) are template-matching techniques most often utilized.
 \item Probabilistic model --- An effective feature vector from pronunciations accurately characterizes the speaker's feature vector's distribution in the feature space. A mathematical model is constructed using statistical characteristics. A few model parameters serve to represent and store mathematical models. The feature vector of the test speech is compared to the mathematical model used to describe the speaker. The similarity between the test speech and the model is computed and helps make the recognition decision. The most widely used model is HMM because HMM correctly captures the properties of human vocal tract alterations and provides a reasonable description of stationarity and variability.
 \item Artificial neural networks (ANN) --- ANN is self-organized and self-learning, which may enhance its performance over time. ANN's features may be utilized to effectively extract speakers' personality traits from audio samples to implement SI systems.
\end{enumerate}

Several performance metrics for evaluating the SI system include recognition rate, training duration, number of training corpora, reaction time, speaker set size, speaking mode, pricing, and number of training corpora. There are several assessment indicators for various events. The recognition rate is the most crucial factor, and it must be assured first to serve as the baseline for all other performance measures. Accurate and false recognition rates are frequently employed in voice recognition systems. The speaker confirmation mechanism determines the erroneous rejection and acceptance rates. The two are at odds with one another. Different sizes are needed for various events. The equal error probability is crucial for assessing speaker confirmation since it states that the two are equivalent if a particular judgment threshold is met.

\subsection{Metrics}
Word error rate (WER) and sentence error rate (SER) are the most often utilized metrics because the voice assistants' primary responsibility is to convert spoken words into text. WER is the proportion of the number of words in the common word order divided by the number of added, changed, or removed words. SER is the number of sentence recognition errors divided by the total number of sentences. However, SER is typically 2 to 3 times higher than WER. Therefore, SER is frequently ignored.

 

\subsection{Taxonomy}

\begin{figure*}[!ht]
\centering
  \includegraphics[width=14cm]{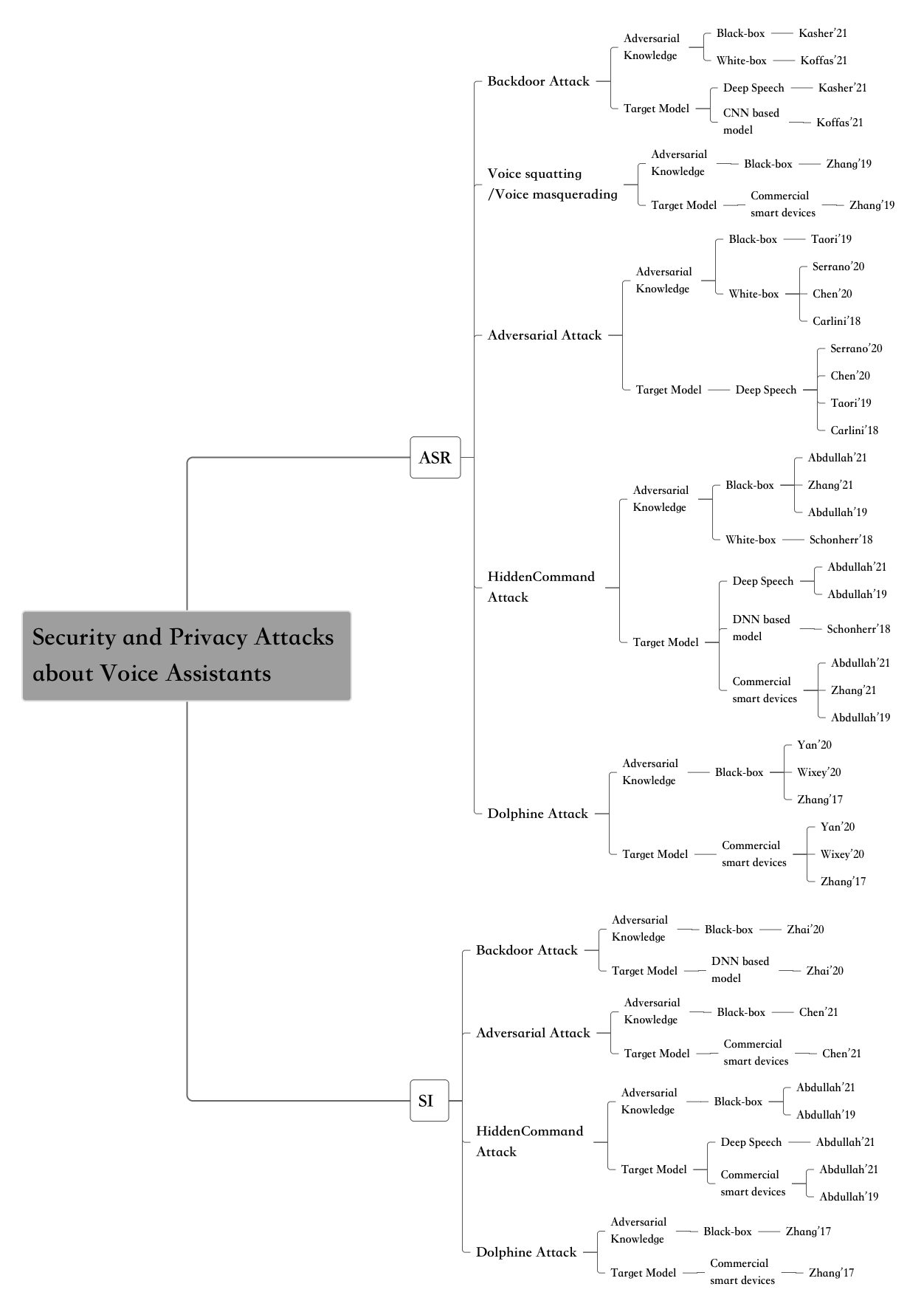}
  \caption{Taxonomy of security and privacy attacks towards voice assistant applications.}
  \label{fig:attVAA}
\end{figure*}

We develop a taxonomy of security and privacy problems in the voice assistant domain to define the attacks against voice assistants. This taxonomy classifies voice assistants' security and privacy risks recently published. Our taxonomy investigates various target models, adversarial information, and attack strategies. We further classify publications in the target model level category based on probabilistic models and target machine learning model types, such as DNN, RNN, CNN, and many alike. Popular apps are also used in various empirical studies, including Amazon Alexa, Google Assistant, and Microsoft Cortana. We categorize articles according to adversarial knowledge levels as black-box, grey-box, and white-box attacks on the target model. The taxonomy of attacks that threaten voice assistant models is shown in Figure \ref{fig:attVAA}. The current attacks on SI models include spoofing, backdoor, adversarial, and hidden command attacks. Attacks in ASR models include dolphin attacks, adversarial attacks, and hidden command attacks. 

\begin{figure*}[!ht]
\centering
  \includegraphics[width=18cm]{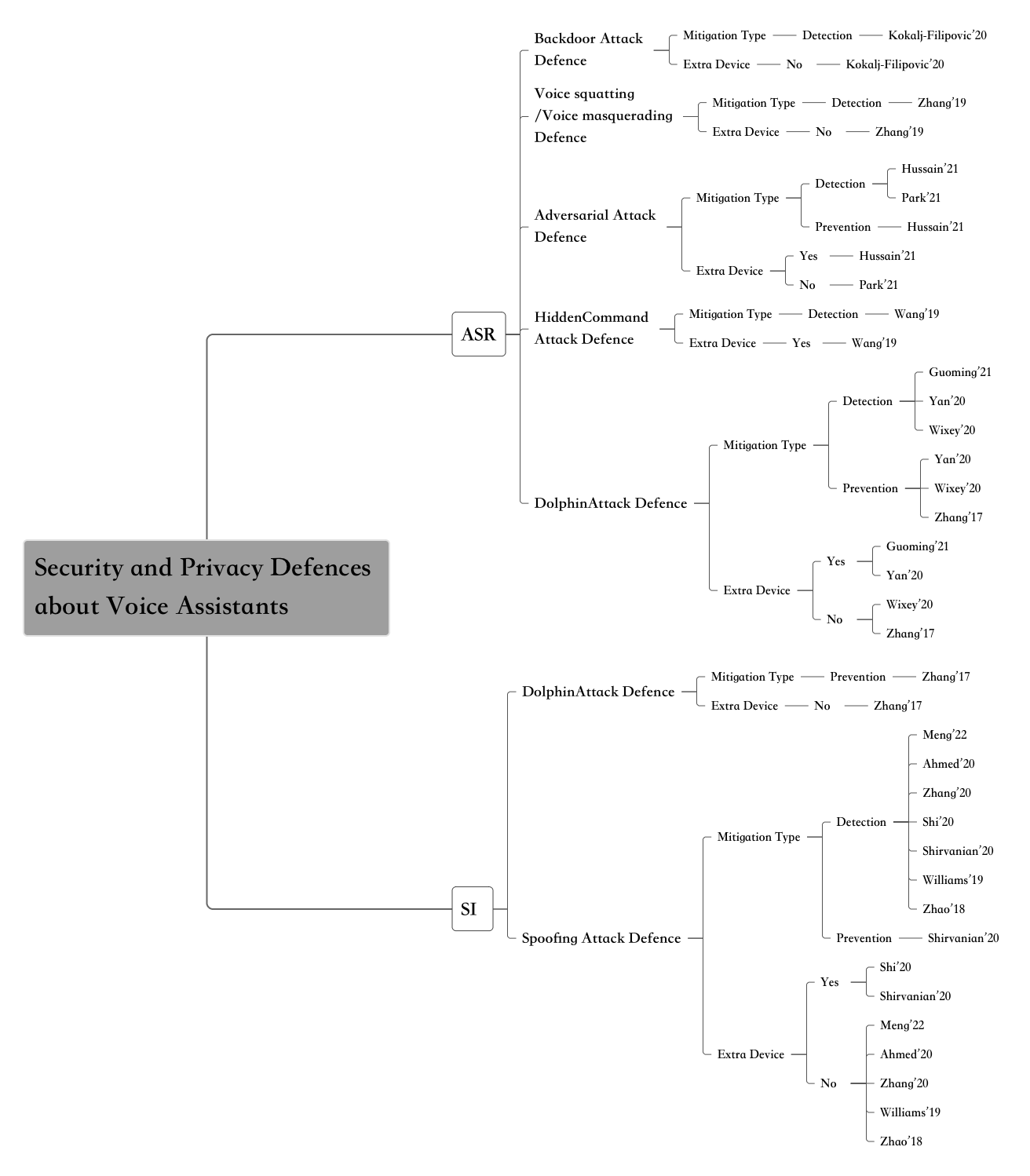}
  \caption{Taxonomy of security and privacy defenses towards voice assistant applications.}
  \label{fig:defVAA}
\end{figure*}

We also developed a taxonomy for defensive methods in the voice assistant domain. The taxonomy of defensive methods is shown in Figure \ref{fig:defVAA}. The defensive methods were listed based on different attacks that they mitigate. Furthermore, each mitigation method was categorized based on its mitigation types: detection and prevention. The defensive methods also were categorized based on whether they needed extra devices when they are deployed to protect the voice assistant applications.

We categorize all published publications that discuss attacks on ASR and SI models and then choose a few exemplary studies to put in Tables 1 and 3. Each chosen paper either revealed new ways to exploit privacy issues on a particular target model or offered new attack strategies. Table 1 and Table 3 provide additional details about each study that in the taxonomy Figure 4, which can aid readers in understanding and comparing each work. We specifically provide the publication year, publication venue, learning task for the target model, attack knowledge available to the attacker, specific attack approach, baseline for the proposed attack, metrics for evaluating attack performance, and datasets used in the experiments for each paper listed in Tables 1 and 3.

\section{Attacks and Defences in ASR-based Voice Assistant Applications}
\label{sec:asr}

\subsection{Backdoor Attacks and Defences in ASR}

\begin{table*}[!ht]
\caption{A comparison of existing attacks on ASR systems. The medium of a practical attack includes: A: Over-the-air, L: Over-the-line, P: Over-the-phone, S: Over-the-surface. The target model of a practical attack includes: \uppercase\expandafter{\romannumeral1})Deep Speech, \uppercase\expandafter{\romannumeral2})CNN based model, \uppercase\expandafter{\romannumeral3}) RNN based model, \uppercase\expandafter{\romannumeral4}) DNN based model, \uppercase\expandafter{\romannumeral5})Commercial smart devices. The setting of a practical attack includes: B: Black-box, W: White-box, G: Grey-box.}
\label{tab:attASR}
\resizebox{\textwidth}{!}{%
\begin{tabular}{|c|c|c|c|c|c|c|c|c|c|}
\hline
\multirow{2}{*}{Type} &
  \multirow{2}{*}{Year} &
  \multirow{2}{*}{Paper} &
  Medium &
  \multirow{2}{*}{Goal} &
  Target Model &
  Setting &
  \multirow{2}{*}{Attack Approach} &
  \multirow{2}{*}{Success} &
  \multirow{2}{*}{Transferability} \\ \cline{4-4} \cline{6-7}
               &      &              & A  L  P  S &   & \uppercase\expandafter{\romannumeral1}  \uppercase\expandafter{\romannumeral2}  \uppercase\expandafter{\romannumeral3} \uppercase\expandafter{\romannumeral4}  \uppercase\expandafter{\romannumeral5}     & B  W  G &                                &                        &    \\ \hline
\multirow{2}{*}{Backdoor}       & \multirow{2}{*}{2021} & Kasher'21    & $\bigstar$ $\square$ $\square$ $\square$    & T & $\bigstar$  $\square$  $\square$  $\square$  $\square$ & $\bigstar$ $\square$ $\square$   & Comparison of Characteristics    & 50\%                   & Y  \\ \cline{3-10}
               &      & Koffas'21    & $\bigstar$ $\square$ $\square$ $\square$    & T & $\square$  $\bigstar$  $\square$  $\square$  $\square$ & $\square$ $\bigstar$ $\bigstar$   & Dataset Poisioning             & \textgreater{}=99\%    & PY \\ \hline
\begin{tabular}[c]{@{}c@{}}Voice squatting\\ /Voice masquerading\end{tabular} &
  2019 &
  Zhang'19 &
  $\bigstar$ $\square$ $\square$ $\square$ &
  T &
  $\square$  $\square$  $\square$  $\square$  $\bigstar$ &
  $\bigstar$ $\square$ $\square$ &
  Use third party market &
  \textgreater{}=50\% &
  - \\ \hline
\multirow{5}{*}{Adversarial}    & \multirow{2}{*}{2020} & Serrano'20   & $\square$ $\bigstar$ $\square$ $\square$    & U & $\bigstar$  $\square$  $\square$  $\square$  $\square$ & $\square$ $\bigstar$ $\square$   & Adversarial perturbations      & -                      & PY \\ \cline{3-10}
               &      & Chen'20      & $\bigstar$ $\square$ $\square$ $\square$    & T & $\bigstar$  $\square$  $\square$  $\square$  $\square$ & $\square$ $\bigstar$ $\square$   & Domain adaptation algorithm    & D, \textgreater{}=90\% & -  \\ \cline{2-10}
 &
  2019 &
  Taori'19 &
  $\square$ $\bigstar$ $\square$ $\square$ &
  T &
  $\bigstar$  $\square$  $\square$  $\square$  $\square$ &
  $\bigstar$ $\square$ $\square$ &
  \begin{tabular}[c]{@{}c@{}}Combine genetic algorithms \\ and gradient estimation\end{tabular} &
  35\% &
  - \\ \cline{2-10}
               & 2018 & Carlini'18   & $\square$ $\bigstar$ $\square$ $\square$    & T & $\bigstar$  $\square$  $\square$  $\square$  $\square$ & $\square$ $\bigstar$ $\square$   & End-to-end                     & 100\%                  & PY \\ \hline
\multirow{4}{*}{Hidden Command} & \multirow{2}{*}{2021} & Abdullah'21  & $\square$ $\square$ $\bigstar$ $\square$    & U & $\bigstar$  $\square$  $\square$  $\square$  $\bigstar$ & $\bigstar$ $\square$ $\square$   & Interfere signal preprocessing & 100\%                  & Y  \\ \cline{3-10}
               &      & Zhang'21     & $\bigstar$ $\bigstar$ $\square$ $\square$    & T & $\square$  $\square$  $\square$  $\square$  $\bigstar$ & $\bigstar$ $\square$ $\square$   & Faste speech                   & D, \textgreater{}=90\% & Y  \\ \cline{2-10}
               & 2018 & Schonherr'18 & $\square$ $\bigstar$ $\square$ $\square$    & T & $\square$  $\square$  $\square$  $\bigstar$  $\square$ & $\square$ $\bigstar$ $\square$   & Psychoacoustic hiding          & 98\%                   & PY \\ \cline{2-10}
               & 2019 & Abdullah'19  & $\bigstar$ $\bigstar$ $\square$ $\square$    & T & $\bigstar$  $\square$  $\square$  $\square$  $\bigstar$ & $\bigstar$ $\square$ $\square$   & Generic attack method          & 80\%                   & Y  \\ \hline
\multirow{2}{*}{DolphinAttack}  & \multirow{2}{*}{2020} & Yan'20       & $\square$ $\square$ $\square$ $\bigstar$    & T & $\square$  $\square$  $\square$  $\square$  $\bigstar$ & $\bigstar$ $\square$ $\square$   & Solid material                 & D, \textgreater{}=90\% & -  \\ \cline{3-10}
               &      & Wixey'20     & $\square$ $\bigstar$ $\square$ $\square$    & U & $\square$  $\square$  $\square$  $\square$  $\bigstar$ & $\bigstar$ $\square$ $\square$   & Internet attack                & -                      & PY \\ \cline{2-10}
 &
  2017 &
  Zhang'17 &
  $\bigstar$ $\square$ $\square$ $\square$ &
  T &
  $\square$  $\square$  $\square$  $\square$  $\bigstar$ &
  $\bigstar$ $\square$ $\square$ &
  Modulate the frequency of the voice signal &
  D, \textgreater{}=90\% &
  PY \\ \hline
\end{tabular}%
}

$\bigstar$ Applicable \quad  $\square$ Not applicable \quad T: Targeted \quad U: Untargeted \quad D: Depending on the experimental environment  \quad  Y: Yes  \quad PY: Presumably Yes \quad - None (or unspecified)
\end{table*}

\begin{table*}
 \caption{A comparison of existing defensive methods for ASR systems. The type of a defence includes: D: Detection, P: Prevention}
 \label{tab:defASR}
\resizebox{\textwidth}{!}{%
\begin{tabular}{|c|c|c|c|c|c|c|c|}
\hline
\multirow{2}{*}{Mitigated Attacks} &
  \multirow{2}{*}{Year} &
  Type &
  \multirow{2}{*}{Paper} &
  \multirow{2}{*}{Methods} &
  \multirow{2}{*}{Extra Device} &
  \multirow{2}{*}{Success} &
  \multirow{2}{*}{Transferability} \\ \cline{3-3}
                               &   &  D   P  &                     &                                                &   &       &    \\ \hline
Backdoor                       &  2020  & $\surd$  $\times$ & Kokalj-Filipovic'20 & Deep Learning Classifiers                      & $\times$ &   -    & Y  \\ \hline
\begin{tabular}[c]{@{}c@{}}Voice squatting\\ /Voice masquerading\end{tabular} &
 2019  &
  $\surd$ $\times$ &
  Zhang'19 &
  Context-sensitive detector &
  $\times$ &
  95\% &
  - \\ \hline
\multirow{2}{*}{Adversarial}   &  \multirow{2}{*}{2021}  & $\surd$ $\surd$ & Hussain'21          & LPC and Mel extraction inversion               & $\surd$ & -     & Y  \\ \cline{3-8} 
                               &     &  $\surd$ $\times$    & Park'21             & Add noise to logits                            & $\times$ & -     & PY \\ \hline
Hidden Command                 &  2019  & $\surd$ $\times$ & Wang'19             & Extract vibration features from motion sensor  & $\surd$ & 100\% & Y  \\ \hline
\multirow{4}{*}{DolphinAttack} & 2021   & $\surd$ $\times$ & Guoming'21          & Decay rate difference                          & $\surd$ & 99\%  & Y  \\ \cline{2-8} 
                               &  \multirow{2}{*}{2020}  & $\surd$ $\surd$ & Yan'20              & Monitor the characteristics of the source signal & $\surd$ & -     & -  \\ \cline{3-8} 
                               &    &   $\surd$ $\surd$   & Wixey'20            & Use imperceptible sounds as covert channels    & $\times$ & -     & Y  \\ \cline{2-8} 
 & 2017
   &
  $\times$ $\surd$ &
  Zhang'17 &
  \begin{tabular}[c]{@{}c@{}}Hardware-based: microphone enhancement \\ and baseband cancellation; Software-based: machine learning\end{tabular} &
  $\times$ &
  - &
  PY \\ \hline
\end{tabular}%
}
$\surd$ Positive \quad $\times$ Negative \quad - None (or unspecified)\quad   Y: Yes \quad PY: Presumably Yes
\end{table*}

Backdoor attacks are common attacks against ASR models. Backdoor systems refer to attackers embedding inaudible signals into data (music clips or voicemail messages), disrupting the results of how that data is processed in a machine-learning model. A backdoor attack involves inserting a particular input of a secret command, often referred to as a trigger, into the model such that the model will make the attacker's desired decision with a specific input. Adversarial audio translates concealed orders into the attacker's intended command using the voice assistant's neural processing network, making hidden commands perceptive to human ears. Kasher et al.~\cite{Kasher2021} used the backdoor system to control several smart devices that allow voice commands, introducing potent, noise-resistant adversarial audio perturbations to typical voice as music-only audio to translate target commands. To maximize the impact of backdoor, our study evaluates various base vectors, target words, and perturbation strengths. Their backdoor method is equally effective in delivering perturbations applied to musical samples as it is used to speech-based samples, enabling the perturbation of various samples. Although the target phrase is important and the orders to be conveyed must be carefully chosen, it is plausible to attain a transcription accuracy rate of more than 50\%.

Many businesses frequently outsource the training process to third parties or purchase pre-trained models to save costs. Unfortunately, outsourcing gives various adversaries access points, such as backdoor attacks. Koffas et al.~\cite{Koffas2021} look into backdoor attacks against ASR systems. Injecting inaudible triggers makes it more difficult to identify a backdoor attack. They employed datasets with 10 and 30 classes along with three neural networks (two CNNs and one LSTM), respectively. They also looked at the impact of trigger type, duration, and location. The findings demonstrate that launching an inaudible Backdoor attack against ASR is simple, requiring only that the attacker poison just around 0.5 percent of the training sample. Because the trigger cannot be heard, it may be as long as the signal duration, which increases the attack's potency. Discontinuous triggers can also dramatically enhance attack performance, even with brief triggers. Short non-consecutive triggers that are put into less than 0.5 percent of the training dataset can result in attack success rates higher than 99 percent.

Though the attack strategy in those papers was quite effective, no defensive mechanism was studied. Therefore, defenses will be a focus of future research in the Backdoor attacks domain. Kokalj-Filipovic et al.~\cite{Kokalj-Filipovic2020} argued for more investigation in detecting adversarial backdoor samples. To examine the detectability of backdoor channels, the raw acoustic data were utilized in covert attacks. Backdoor channels contain inaudible messages using 40kHz sine waves for modulation and ultrasonic speakers for transmission. To analyze the neural networks, a backdoor dataset is created using the incoming composite signal. The sound samples served as the foundational dataset for training and were played back and recorded. The backdoor dataset is used to assess the backdoor channel's effects on the categorization of acoustic data, and trials reveal that the classifier's accuracy declines. The deterioration varies by deep classifier type and appears to be less pronounced for those trained with autoencoders. They suggest statistics, such as the empirical entropy layer of the classifier output or the log-likelihood of a variational autoencoder used to pre-train a classifier, that may be used to identify out-of-distribution data generated by the backdoor channel. The first findings in their work indicate that more research is necessary before using deep learning classifiers such as Backdoor covert channel detectors.

In summary, a backdoor attack refers to hiding the inaudible signals in data and go in into the model through a backdoor channel to affect the transcript of the input data. Backdoor attacks are not limited to human voices, and the samples can be pure music clips or noisy audio that does not result in transcription without the perturbations. Moreover, the trigger signal might last as long as the injected audio. The prolonged trigger makes the attack more powerful without exposing the signal. The attack rate of the recent Backdoor attack technique can reach over 90\% on average.

\subsection{Voice Squatting and Voice Masquerading}
Voice assistants are a ubiquitous part of people's daily lives in the modern world. One must first authenticate to hear instructions and call other services using a voice assistant. As the third-party market grows, more and more service providers add new functions (skills) to it. As a result, attackers can trick voice assistants by uploading malicious skills to unofficial markets.

To evaluate how much impact the skills in the third-party market have on the security of voice applications, Zhang et al.~\cite{Zhang2019} examine whether such remote, massive attacks through the third-party market are plausible. Voice squatting and masquerading were identified as two novel attack techniques. Voice squatting means imitating a legal ability by employing a wake phrase that sounds like the legal skill you intend to use. For instance, employing the lawful skill ``open capital one" prevents voice assistants from utilizing the criminal skill ``capital won". When you hear ``capital one please," call another talent. When an attacker uses speech masquerading to communicate with a user and get sensitive information, they pose as a voice assistant or a called skill. The trials the authors did on the Google Home and Amazon Echo, which included user research and actual installations, revealed that they both offer a genuine threat.

Amazon and Google constructed a new squat detector available on Alexa and Google's marketplace. It confirms the importance of voice squatting and masquerading. A skill named scanner was created and available on the Amazon and Google Skills marketplaces to mitigate the risk. It identified several vulnerable Alexa skills and published skill names, indicating that the attack may have affected thousands of VPA customers. Additionally, they developed and used a context-sensitive detector to 95\% accurately reduce the threat of voice masquerading. Future work will be needed to improve voice channel security and to verify the people engaged without interfering with the VPA system's availability.

\subsection{Adversarial Attacks and Defences in ASR}

Recurrent neural networks are widely employed in time series data machine learning systems, especially in ASR systems. Serrano et al.~\cite{Serrano2020} suggest that RNNs are susceptible to periodic adversarial perturbations because of their distinctive memory and parameter-sharing capabilities. They create an attack to produce antagonistic perturbations in the DeepSpeech model, resulting in a collection of previously unknown cases generated in real time. According to the experimental findings, there is an average 0.426 cross-correlation between each undisturbed audio and each disturbed audio sample, and an average 0.017 mean square error between the two. Between the original, unaltered DeepSpeech transcript and each disturbed DeepSpeech transcript, the smoothed BLEU score is 0.221. Although there have been other white-box and black-box iterative attacks on RNNs, their strategy takes advantage of the problem's temporal nature. RNNs have the memory to schedule and generate adversarial perturbations in real-time using deep reinforcement learning, making it more practical.

Carlini et al.~\cite{carlini2018audio} launched an end-to-end adversarial attack against the DeepSpeech model in a white box environment. Although optimizing with MFC preprocessing procedures is difficult, they directly fed the unprocessed data as the classifier's input. Regardless of the necessary transcription or source audio samples, their attack was 100\% effective. They employed samples that begin with arbitrary waveforms, like music clips, to embed speech within audio that should not be identified as speech. They have the potential ability to transcribe speech into music, conceal speech from being transcribed, and transcribe audio at a rate of up to 50 characters per second. It reveals the value of focused audio adversarial samples for automated speech recognition. Hence, adversarial audio samples vary from sampling on pictures by demonstrating that linearity does not hold in the audio domain.

Adversarial attacks on deep recurrent networks continue to improve in ASR systems. Most earlier research was conducted in a ``white box" context, which required model design and parameter knowledge. Taori et al.~\cite{taori2019targeted} use a combination of evolutionary algorithms and gradient estimation techniques to generate adversarial samples in a black box environment. In tests, they maintained 94.6 percent audio file similarity while achieving 89.25 percent target attack similarity and a 35 percent target attack success rate after 3,000 generations. According to experimental findings, the adversarial samples produced by combining the genetic algorithm with gradient estimation are superior to those produced by either approach alone. The method can achieve perfect or nearly perfect goals on most audio sample transcription while maintaining a high degree of similarity. The genetic algorithms were used to explore more space by encouraging random mutation and ending with a more guided search of gradient estimates. This straightforward technique performs targeted adversarial attacks on black-box models. The method's efficiency is increased by adding momentum mutations and noise designed particularly for high frequencies. By avoiding noise from obstructing human voices and objectively enhancing the experimental audio sample correlation, limiting noise to the high-frequency domain increases their similarity. The production of adversarial instances produces the best results when all four techniques are combined. In particular, this research targets deep nonlinear ASR systems that may generate translations of any length and provides a new category of black-box attacks.

Reducing the additional disruption caused by signal distortion during propagation is a major challenge in aerial transmission. Chen et al.~\cite{Chen2020Metamorph} proposed producing inaudible sounds so that the generated signal can withstand air transmission. The frequency selectivity of devices and channels is the leading cause of signal distortion. Based on this characteristic, Metamorph employs a ``create and clean" two-phase architecture. A few prior measurements were used to create an initial perturbation that solely accounts for the impacts of core distortion before a domain adaptation technique was used to increase the attack range and reliability. At attack ranges as close as 6 meters, the assessment recorded a high attack success rate of 90\%.

The previous introduction shows that deep learning-based ASR systems face severe security and privacy problems, particularly adversarial attacks, due to the rapid development of deep learning and the widespread use of voice assistants. According to most studies, the attack's success rate is over 90\% on average. Hussain et al.~\cite{hussain2021waveguard} investigated the identification of adversarial cases and the response against adversarial attacks. WaveGuard is a framework for spotting adversarial inputs that aim to undermine ASR systems. WaveGuard combines audio transformation capabilities to identify adversarial inputs and evaluates ASR transcriptions of raw and converted audio. WaveGuard can detect adversarial instances from four current audio adversarial attacks with various audio transition algorithms.

The robustness of defense technology is vital against adversarial attacks. Hussain et al.~\cite{hussain2021waveguard} examined defense's robustness in the audio domain regarding survival adaptation and defense evaluation. The adaptive adversary could only avoid the transformation that compresses the input into an audio representation of perceptual information if the adversarial perturbation's size was large. In contrast, a naïve audio conversion function can be readily defeated by an adaptive attacker with a negligible and inaudible disturbance. As a result, transformations like LPC and Mel extraction inversion become potent options for blocking adversarial audio attacks. Furthermore, even in fully white-box environments, audio transformations that recover audio from perceptually informed representations can strengthen defenses against adaptive adversaries. Additionally, WaveGuard can be quickly and easily incorporated with any ASR model without retraining it to recognize audio adversarial cases effectively. Park et al.~\cite{park2021detecting} proposed to identify audio adversarial cases. Before supplying logits to the ASR's decoder, they selected noise to impact the transcription results of audio adversarial cases without impact on the transcription results of benign audio waves. They tested this technique on DeepSpeech to gauge the effectiveness of their suggested approach. They assessed various tactics, including two airborne and three internet attacks. This method has greater accuracy than current state-of-the-art techniques. The approach is straightforward, does not require model retraining, and is resilient against online and aerial audio adversarial samples.


\subsection{Hidden Command Attacks and Defences in ASR}
The deep neural network (DNN) helps ASR match human hearing and understanding in many tasks. Unfortunately, DNNs are highly vulnerable to adversarial perturbations. A hidden command attack is more advanced than the common adversarial attack introduced in the last section. Because the perturbations are imperceptible by human ears, hidden orders are noise signals that ASR correctly understands while humans do not.

Schonherr et al.~\cite{schonherr2018adversarial} present an innovative adversarial example based on psychoacoustic concealment. Their attack uses DNN-based ASR systems' characteristics, adding a back-propagation stage to the initial analytic process. They tested the cutting-edge speech recognition system Kaldi to identify the optimal performance parameters and settings for various inputs. For a 10-second audio recording, the attack is successful in 98 percent of instances with a computing effort of fewer than two minutes. It was discovered that human listeners could not understand their goal transcription. Although the accuracy remained unchanged, the users could still understand the original voice content.

The DNN-HMM system is subjected to forced alignment and backpropagation, which produces undetectable adversarial perturbations with high reliability. Schonherr et al.~\cite{schonherr2018adversarial} assessed various algorithm variables, such as the number of iterations and permitted hearing threshold deviations. Their strategy involves significantly less distortion than approaches suggested by other studies, which can provide focused adversarial situations. Future research may strengthen ASR systems by considering psychoacoustic models to thwart these simple attacks. Additionally, black-box settings for real-world attacks and commercial ASR systems should be used to evaluate comparable attacks.

The concealed command injection worked well. White-box settings have been the focus of earlier research on forced mistranscription or misrecognition, utilizing white-box information to carry out targeted attacks on a model. However, those attacks frequently rely heavily on a white-box understanding of specific machine-learning models and are restricted to a small number of microphones and speakers, limiting their applicability across various acoustic hardware platforms. Abdullah et al.~\cite{Abdullah'19} utilize a model-agnostic (black-box) approach that takes advantage of understanding signal processing methods frequently used by VPS to produce data for input into machine learning systems to eliminate these dependencies and make hidden command attacks more practicable. Different source audio samples have similar feature vectors when converted by an acoustic feature extraction method like FFT. 12 machine learning models, including 7 proprietary models (such as Google Speech API, Bing Speech API, IBM Speech API, Azure Speaker API, etc.) were evaluated and presented against all targets. They also established four classes of perturbations that result in unintelligible sounds. Furthermore, they demonstrated efficacy in both model and actual systems by using fraudulently produced audio samples in various hardware setups.

The use of domain-specific knowledge is an effective method for creating successful covert voice command attacks. Abdullah et al.~\cite{Abdullah'19} study attack audio based on feature vectors produced by signal processing methods rather than attacking the underlying machine learning model. Using a black-box method, their technique is appropriate for a range of Over-the-Wire and Over-the-Air speech detection and SI systems. In 2021, Abdullah et al.~\cite{Abdullah'21} investigated the pipeline phase as a black box and transferrable attack. By changing just a few audio frames, the new attack reached a 100\% mistranscription and misrecognition rate. Models may pick up on minor, human-imperceptible components of speech that are crucial to the model's accuracy. It explains why eliminating certain inconsequential speech fragments may confuse the model and result in inaccurate transcription or misrecognition. Additionally, some English phonemes, particularly vowels, were much more susceptible. These attacks operate well on cellular networks where transcoding, jitter, and packet loss weaken the signal. These unperceivable phrases or signals that can significantly affect the accuracy of ASR or SI systems will be used to construct more potent models that can explain and lower ASR and AVI system faults. Additionally, they discovered that while adversarial training could aid in partial mitigation, stronger defenses are eventually needed to ward off these attacks.

In the continued path of discovering new hidden command attack methods, most researchers rely on white-box knowledge or a restricted transmission environment. Zhang et al.~\cite{Zhang'21CommanderGabble} suggested an easier-to-build attack independent of the target model. Rapid speech masks voice commands since people and ASR systems frequently mistake fast speech. Specifically, it is possible to persuade ASR to infer secret meanings from the high-speed version of the manipulation by carefully changing the phonetic structure of the target voice command. Their tests were against seven real-world ASR systems. The wireless attack consistently succeeds for all 18 chosen instructions in various real-world settings so that it can pass off up to 96 of the 100 voice requests put to the test as adversarial commands. Their research reveals how poorly current ASR systems handle fast speech and successfully make adversarial audio by fusing fast speech with phoneme manipulation. Additionally, they empirically illustrate CommanderGabble's reliability and usefulness in several real-world situations.

Because defending hidden command attacks is difficult, there are fewer defense research outcomes than attack research. Many mobile gadgets and voice assistant systems employ voice access technologies, which are frequent targets of concealed command injection attacks. Attorneys incorporate adversarial disturbances to make voice assistants provide fraudulent orders without being observed by legitimate users. The detection technique suggested in Wang et al.~\cite{Wang'19} entails identifying and analyzing the distinctive audio signatures of spoken instructions sent in the vibration domain. Wang et al.~\cite{Wang'19} distinctive audio-induced surface vibrations detected by motion sensors are difficult to replicate, and the carefully designed audio signatures concealing verbal orders could trick authentication systems through audio. Assailants will have a complex problem since it requires them to simulate acoustic fingerprints in both audio and vibration domains. They employ a learning-based method to distinguish between conventional and covert voice instructions by extracting temporal and frequency-domain statistical characteristics and acoustic signals (such as chroma vectors and MFCCs) from motion sensor data. According to experimental data, their system can distinguish between disguised and conventional voice instructions with 99.9\% accuracy by employing a low-cost motion sensor with a very low sample frequency. Additionally, the system gains from a speaker motion sensor.

However, Wang et al.~\cite{Wang'19} system's playback procedure might result in front-end mode incursion and playback delays. While n-times speed playback and partial command playback can attain excellent accuracy, it expedites the playback process to 0.5 seconds. To achieve zero incursion, a potential option is to modulate the front-end playback sound to inaudible frequencies (for example, higher than 16KHz). While playing the front-end playback sound, many studies use a rear-end playback configuration with hidden and inexpensive devices with built-in speakers and motion sensors. Additionally, their unsupervised learning-based strategy may be expanded to protect against additional threats, such as ultrasonic strikes, without requiring much training. It is also worthwhile to investigate if playback noises and human voices can be separated in the vibration domain.

\subsection{DolphinAttack and its Defences in ASR}
Speech assistants are susceptible to transcribed signal injection at inaudible frequencies. In earlier studies, high-frequency signal injection into audio has received much attention. Sound waves above or below the human hearing range are used as an attack strategy in DolphinAttack. There are many ways to attack voice assistants, and research has concentrated on inserting secret orders into audio samples. Zhang et al.~\cite{Zhang2017Dolphin} is the first to propose a silent attack named DolphinAttack. Although DolphinAttack uses an ultrasonic carrier wave (with a frequency over 20 kHz) to avoid human perception, voice assistants likely discern the signal because of the hardware properties of the microphone. DolphinAttack was successfully conducted on several voice recognition systems, including Siri, Google Now, Samsung S Voice, Huawei HiVoice, Cortana, and Alexa.

Zhang et al.~\cite{Zhang2017Dolphin} suggested baseband cancellation and microphone augmentation as two hardware-based defense tactics. In terms of microphone improvement, an improved microphone suppresses acoustic sounds with ultrasonic-range frequencies. Signals in the ultrasonic frequency range with AM modulation properties can be identified and demodulated to obtain baseband in terms of baseband cancellation. Since there will be no association between the acquired audible speech signal and the noise in the ultrasonic region, a command cancellation procedure will not affect the microphone's working condition.

However, it is challenging to stop DolphinAttack. To launch DolphinAttack, attackers surreptitiously insert malicious orders into voice assistants and control systems by modulating audible speech with ultrasonic waves (such as doors or smart speakers). Guoming et al.~\cite{Guoming2021EarArray} proposed EarArray as a simple technique for detecting DolphinAttack. EarArray exploits the fact that ultrasonic waves decay more quickly than audible sound to identify DolphinAttack. EarArray compares the command sound signal with the decay rate using numerous microphones integrated into the smart device. EarArray could determine the attacker's direction with 97.89\% accuracy and detect inaudible spoken orders with 99.0\% accuracy. 

The viability of transmitting silent ultrasonic attacks using solid materials is proposed as SurfingAttack in Yan et al.~\cite{Yan2020SurfingAttack}. Unlike wireless transmission in previous studies, the new attack may conceal itself inside or beneath the solid material, opening up a new path for inaudible attacks. SurfingAttack employs the energy transmission mechanism of ultrasonic guided waves, which is a viable and affordable attack method. SurfingAttack effectively attacked devices from 30 feet away while using just 0.75W of attack signal power. Several trials were tested to determine the scope and boundaries of this hazard. The voice response was listened to at a low volume to facilitate dialogue between an opponent and the voice-controllable device. SurfingAttack may allow attackers to decrypt SMS passwords or place phony calls.

Yan et al.~\cite{Yan2020SurfingAttack} proposed several strategies for defense against SurfingAttack. There are three recommendations to reduce or eliminate acoustic vibrations in the ultrasonic range, including improving the microphone's hardware arrangement, positioning the gadget atop a soft woven fabric, and differentiating the frequency difference between attacks and normal signals. 54 attacks were manifested using various attack settings (i.e., frequency, table material, distance, baseband signal, and device). Because the human voice contains few extremely high-frequency components, the received voice signal will be labeled an attack if the attack index is higher than the pre-set threshold. This defense technique may fail if the device's audio low-pass filter has a cut-off frequency lower than 10 kHz because of inadequate data for attack index calculation.

The network attacks in Wixey et al.~\cite{Wixey2020} cause smart devices to create high-frequency (17-21 kHz) independently and low-frequency (60-100 Hz) sounds, transforming them into acoustic attack weapons. Several gadgets appeared capable of emitting noises at volumes that matched or surpassed several advised limits. The measurements were conducted in an anechoic chamber at a distance of one meter. It can impact individuals across a wide region and be used for large and lethal devices. For instance, a speaker system in a car or a linked PA system during a concert or sporting event may be attacked to emit dangerous noise to human beings. Other, ``noisier" channels may also be used, including smart TV broadcasts and injecting HFN or LFN into phone conversations.

Wixey et al.~\cite{Wixey2020} evaluated two free Android applications with an external microphone to generate an alarm when the sound intensity increased, particularly HFN. Other defense tactics against Wixey's attack include routing inaudible noises, restricting speakers' frequency range, notifying the user, disabling playing audio files outside the audible range, and strengthening mobile app permissions. Consumer and business antivirus detection engines can incorporate heuristics to find these threats. A confirmation prompt may be sent to the user to ask if they want to proceed if, for instance, a combination of specific behaviors is frequently identified by antivirus engines as suspicious behavior.

\begin{table*}[!ht]
\caption{A comparison of existing attacks on SI systems. The medium of a practical attack includes: A: Over-the-air, L: Over-the-line, P: Over-the-phone, S: Over-the-surface. The goal of a practical attack includes: T: Targeted, U: Untargeted. The target model of a practical attack includes: \uppercase\expandafter{\romannumeral1})Deep Speech , \uppercase\expandafter{\romannumeral2})CNN based model , \uppercase\expandafter{\romannumeral3}) RNN based model , \uppercase\expandafter{\romannumeral4}) DNN based model, \uppercase\expandafter{\romannumeral5})Commercial smart devices. The setting of a practical attack includes: B: Black-box, W: White-box, G: Grey-box.}
\label{tab:attSI}
\centering
\resizebox{\textwidth}{!}{%
\begin{tabular}{|c|c|c|c|c|c|c|c|c|c|}
\hline
\multirow{2}{*}{Type} &
  \multirow{2}{*}{Year} &
  \multirow{2}{*}{Paper} &
  Medium &
  Goal &
  Target Model &
  Setting &
  \multirow{2}{*}{Attack Approach} &
  \multirow{2}{*}{Success} &
  \multirow{2}{*}{Transferability} \\ \cline{4-7}
               &      &             & A  L  P  S & T   U & \uppercase\expandafter{\romannumeral1}  \uppercase\expandafter{\romannumeral2}  \uppercase\expandafter{\romannumeral3} \uppercase\expandafter{\romannumeral4}  \uppercase\expandafter{\romannumeral5} & B  W  G &                                            &       &    \\ \hline
Adversarial    & 2021 & Chen'21     & $\bigstar$ $\bigstar$ $\square$ $\square$    & $\bigstar$ $\bigstar$   & $\square$ $\square$ $\square$ $\square$ $\bigstar$ & $\bigstar$ $\square$ $\square$   & Optimization                               & 99\%  & PY \\ \hline
Backdoor       & 2020 & Zhai'20     & $\square$ $\bigstar$ $\square$ $\square$    & $\bigstar$ $\square$   & $\square$ $\square$ $\square$ $\bigstar$ $\square$ & $\bigstar$ $\square$ $\square$   & Clustering-based attack                    & 45\%  & Y  \\ \hline
hidden command & 2021 & Abdullah'21 & $\square$ $\square$ $\bigstar$ $\square$    & $\square$ $\bigstar$   & $\bigstar$ $\square$ $\square$ $\square$ $\bigstar$ & $\bigstar$ $\square$ $\square$   & Interfere signal preprocessing             & 100\% & Y  \\ \hline
hidden command & 2019 & Abdullah'19 & $\bigstar$ $\bigstar$ $\square$ $\square$    & $\bigstar$ $\square$   & $\square$ $\square$ $\square$ $\square$ $\bigstar$ & $\bigstar$ $\square$ $\square$   & Generic attack method                      & 80\%  & Y  \\ \hline
DolphinAttack  & 2017 & Zhang'17    & $\bigstar$ $\square$ $\square$ $\square$    & $\bigstar$ $\square$   & $\square$ $\square$ $\square$ $\square$ $\bigstar$ & $\bigstar$ $\square$ $\square$   & Modulate the frequency of the voice signal & D     & PY \\ \hline
\end{tabular}%
}
$\bigstar$ Applicable \quad $\square$ Not applicable \quad   D: Depending on the experimental environment \quad   Y: Yes \quad   PY: Presumably yes
\end{table*}

\begin{table*}[!ht]
\caption{A comparison of existing defensive methods for SI systems. The type of a defence includes: D: Detection, P: Prevention}
\label{tab:defSI}
\centering
\resizebox{\textwidth}{!}{%
\begin{tabular}{|c|c|c|c|c|c|c|c|}
\hline
\multirow{2}{*}{Mitigated Attacks} &
  \multirow{2}{*}{Year} &
  Type &
  \multirow{2}{*}{Paper} &
  \multirow{2}{*}{Methods} &
  \multirow{2}{*}{Extra Device} &
  \multirow{2}{*}{Success} &
  \multirow{2}{*}{Transferability} \\ \cline{3-3}
 &
   &
  P  D &
   &
   &
   &
   &
   \\ \hline
\multirow{7}{*}{Spoofing} &
  2022 &
  $\times$ $\surd$ &
  Meng'22 &
  Array Fingerprint &
  $\times$ &
  99.84\% &
  - \\ \cline{2-8} 
 &
  \multirow{4}{*}{2020} &
  $\times$ $\surd$ &
  Ahmed'20 &
  \begin{tabular}[c]{@{}c@{}}Compare the spectral power difference between real human speech \\ and the replayed speech played through speakers\end{tabular} &
  $\times$ &
  91.30\% &
  PY \\ \cline{3-8} 
 &
   &
  $\times$ $\surd$ &
  Zhang'20 &
  \begin{tabular}[c]{@{}c@{}}Catch the dissimilarities between bone-conducted vibrations \\ and air-conducted voices when human speaks\end{tabular} &
  $\times$ &
  97\% &
  N \\ \cline{3-8} 
 &
   &
  $\times$ $\surd$ &
  Shi'20 &
  Cross-domain comparisons: Audio\&vibration &
  $\surd$ &
  97.20\% &
  Y \\ \cline{3-8} 
 &
   &
  $\surd$ $\surd$ &
  Shirvanian'20 &
  \begin{tabular}[c]{@{}c@{}}The WER for the synthesized voices are 2-3 \\ times more than the WER for natural voices\end{tabular} &
  $\surd$ &
  \textgreater{}=95\% &
  - \\ \cline{2-8} 
 &
  2019 &
  $\times$ $\surd$ &
  Williams'19 &
  Combine x-vector attack embeddings with signal processing features &
  $\times$ &
  - &
  - \\ \cline{2-8} 
 &
  2018 &
  $\times$ $\surd$ &
  Zhao'18 &
  Weighting framework with a Gaussian Mixture Model (GMM) classifier &
  $\times$ &
  \textgreater{}=98\% &
  PY \\ \hline
  DolphinAttack &
  2017 &
  $\surd$ $\times$ &
  Zhang'17 &
  \begin{tabular}[c]{@{}c@{}}Hardware-based: microphone enhancement and \\ baseband cancellation; Software-based: machine learning\end{tabular} &
  $\times$ &
  - &
  PY \\ \hline
\end{tabular}%
}
$\surd$ Positive \quad $\times$ Negative \quad - None (or unspecified) \quad   Y: Yes \quad PY: Presumably Yes
\end{table*}

\section{Attacks and Defences in SI-based Voice Assistant Applications}
\label{sec:si}

Most products combine ASR and SI models in the voice assistant applications market. Some applications only provide speech recognition functions. However, stand-alone SI function applications are rare. The research in this area has fewer works that target SI systems only. The number of attack kinds used on SI systems is also fewer than in ASR systems. This section introduces attack and defense techniques for SI systems.

\subsection{Attacks against SI-based Models}
IoT devices often employ the SI system as a biometric identification and authentication method. However, many studies have revealed the system's susceptibility to adversarial attacks. The adversarial attack on the SI system is an open research problem in the black-box context, although the present research generally positively impacts the white-box setting.

In 2021, Chen et al.~\cite{Chen2021Bob} identified flaws in SI systems in black-box environments. It is the first in-depth and systematic research of adversarial attacks against SI systems. To create adversarial samples, the authors suggest using an adversarial technique called FAKEBOB. They develop adversarial instances through an optimization method, combining adversarial example confidence and maximal distortion to balance adversarial sound intensity and imperceptibility. The proposal of a novel technique to estimate the score threshold, a component of SI systems, and utilize it to resolve optimization issues, is a significant contribution of this research. FAKEBOB was tested in 16 attack scenarios with a success rate of 98\%. FAKEBOB could be deployed on Microsoft Azure. Human studies showed that people had trouble distinguishing between neutral and aggressive speakers. FAKEBOB renders useless four potential countermeasures against adversarial attacks in the speech recognition arena.

The training data for many SI systems is collected from third parties, such as Internet open-source databases and third-party data providers. However, using data from unreliable sources frequently puts the SI system's security at risk. Backdoor attacks against SI systems are a security risk when an attacker can poison the training data to introduce a secret backdoor into the speaker verification model. Because those techniques used a single trigger directly and successfully for all poisoned samples, none of the prior backdoor attacks have been able to target SI systems. Zhai et al.~\cite{Zhai2020} planned a cluster-based attack strategy so that poisoned samples from various clusters would have various triggers. The infected model can validate this technique successfully after receiving unregistered defined triggers. Evaluation speaker verification techniques were tested across all datasets in experiments with Attack Success Rates (ASR) greater than or equal to 45\%. The attack is exceedingly difficult to identify since the tainted data's equal error rate (EER) is equivalent to a model built with a good training set. Multiple users' verification data will be present in a SI system in actual applications, increasing the resulting ASR. This approach offers a fresh viewpoint for creating new attacks and a solid foundation for enhancing the robustness of verification techniques.

\subsection{Defensive Methods for SI-based Models}
Speech assistant systems are susceptible to acoustic attacks, including spoofing attacks, because voice signals are accepted in open spaces and channels. Modern VUIs that employ classic voice authentication techniques are susceptible to spoofing attacks, in which an evil adversary impersonates a real user by speaking commands that have already been recorded or synthesized. Once the voice assistant's SI system has been tricked, some risky actions, like making bulk purchases and phoning friends and family, may be executed.

Many studies aim to mitigate this issue to increase the SI system's resistance against spoofing attacks. Constant-Q cepstral coefficients (CQCC) and scattered cepstral coefficients (SCC) are particularly effective for speech synthesis (SS) and speech conversion to identify fake speech signals (VC). However, the equal error rate (EER) cannot be kept low for the subset of lies produced by various approaches. The issue of selective detection degradation is addressed by the proposal of an independent detector with adaptive weighting. To assess data structure, the detector combines a brand-new adaptive weighting framework with a brand-new clustering technique \cite{Zhao2018}. The original CQCC and SCC features are combined at the score level to address the deterioration of a single speech subset. Examine the clustering characteristics of the truthful and dishonest subgroups to select the correct weighting variables. Zhao et al.~\cite{Zhao2018} suggested a weighted framework using a Gaussian Mixture Model (GMM) classifier rather than CQCC or SCC alone because the backend has been found to produce a lower overall EER on the ASVspoof 2015 database. On known and unknown attacks, EERs of 0.01 percent and 0.20 percent, respectively, were recorded. 


Williams et al.~\cite{Williams2019} proposed a novel detection mechanism to determine whether a certain recording is ``genuine" and to research countermeasures against spoofing attacks. The system combines x-vector attack embeddings with signal processing characteristics and employs convolutional neural networks (CNNs) and spoken audio representations. Using a Time Delay Neural Network (TDNN), the x-vectors at the tack embeddings are produced using Mel Frequency Cepstral Coefficients (MFCCs). These embeddings use tagged data to simulate 9 different attack types and 27 different contexts jointly. Additionally, they used SCMCs or sub-band spectral centroid magnitude coefficients. During training, the data was augmented with an additive Gaussian noise layer, increasing the system's resistance to attacks that had not been encountered before. They use the Tandem Detection Cost Function (tDCF) and Equal Error Rate to report system performance (EER). The x-vector attack embeddings can assist in normalizing CNN predictions. Their study employs frame-level 40-dim MFCC features without restricting the frequency range. The best frequency range and frame-level characteristics to employ with these x-vector embeddings to capture variations in utterance acoustic settings could be investigated further.

The previous work introduced useful parameters to achieve a high detection rate, while the following research focused on bringing the mitigation method into practical use. Combine them into hardware that usually sticks with the voice assistant application products or make the mitigation algorithm into software that is easy and safe to download. Shirvanian et al.~\cite{Shirvanian2020Voicefox} provided a simple mitigation strategy without additional hardware equipment to increase the security of speaker verification from voice synthesis attacks. Speech transcribers are essential to their defense. Their method works well because speech-to-text systems interpret synthetic speech differently from authentic voices. The WER for synthetic voices is often 2-3 times more than the WER for genuine voices. Shirvanian et al.~\cite{Shirvanian2020Voicefox} recommended discarding terms not found in the reference dictionary and accepting transcribed text if up to a specific number of words are incorrectly transcribed. Following the application and transcriber being utilized, their detection technology may reduce the false rejection rates to as low as 1.20-4.69 percent and the false accept rates to 0 percent for dictionaries containing phonetically unique terms. Its drawback is that it will not work if a more sophisticated speech synthesis technology can make transcription accuracy equal to that of natural audio. Shi et al.~\cite{Shi2020WearID}  presented a training-free voice authentication system that uses cross-domain voice similarity between the audio and vibration domains to provide increased security for expanding VA system deployments to secure speaker verification systems for voice assistants. Shi et al.~\cite{Shi2020WearID} refers to the system as WearID. When a user provides key instructions, WearID authenticates them by comparing the cross-domain similarity of the distinct speech characteristics recorded by the wearable accelerometer and the VA system's microphone, respectively.

Compared to other security systems mainly used to protect SI systems from spoofing attacks, WearID in \cite{Shi2020WearID} is more efficient in time and labor than previous approaches. WearID does not require the user's active participation (such as responding to messages or phone calls) or the storage of the user's privacy-sensitive voice samples for training. WearID uses a special vibration-sensing interface with a limited audible range (e.g., 25 cm) to validate voice orders. It is difficult to determine whether two sets of domain data are identical. It is challenging to directly compare data from the audio and vibration domains due to the enormous sample rate disparity (e.g., 8000Hz vs.~200Hz), and even little data noise can be magnified and lead to authentication failures. They looked into the intricate connections between the two sensing domains to overcome these difficulties and created a spectrogram-based method that transforms microphone data into low-frequency ``motion sensor data" to enable domain comparisons. They design a user authentication system that uses voice similarities across domains to confirm that voice commands come from authorized users. Numerous tests using two common smartwatches and 1,000 voice commands reveal that WearID can reliably identify users' voice commands with 99.8\% accuracy in regular use and can identify fraudulent voice commands with 97.2\% accuracy in audible/inaudible attacks.

A continuous liveness detection technique for safe VUIs in IoT contexts was developed by Zhang et al.~\cite{Zhang2020VibLive}. It identifies differences between air-conducted vocals and bone-conducted vibrations when a person talks. The detection system validates real users and identifies spoofing attempts without needing users to enroll in particular passphrases. It is a text-independent system without additional software or hardware beyond the standard loudspeaker and microphone found on VUIs. A 97 percent detection accuracy was achieved on 25 participants. Additionally, it uses the weaknesses of most VUIs as a detecting method and needs ultrasound at a frequency of 20 kHz for the probe signal. Hence, this technique exposes the VUIs to an additional risk of DolphinAttack to identify spoofing attempts using replay.

A replay attack is a spoofing attack against voice assistants. Ahmed et al.~\cite{Ahmed2020Void} proposed an effective voice live detection system named ``Void" to mitigate replay attacks. Void compares the spectral power differential between actual human speech and replayed speech played over speakers to identify spoofing attacks. Void utilizes a single classification model with 97 features as opposed to conventional approaches that employ several deep learning models and thousands of data. Two datasets were used to assess the performance: (1) 255,173 speech samples produced by 120 participants, 15 playback devices, and 12 recording devices; and (2) 18,030 publicly available speech sample devices produced by 42 participants, 26 playback devices, and 25 recordings. For each dataset, Void detects voice replay attacks with an error rate of 0.3\% and 11.6\%, respectively. A deep learning-based detection strategy that could lower the detection error rate to 7.4\% was the best-performing detection scheme in the same dataset. However, Void consumes 153 times less memory and eight times faster detection. By combining Void with prior research, a Gaussian mixture model employing Mel Frequency Cepstral Coefficients (MFCCs) as classification features may obtain an error rate of 8.7\%. Void works on a single effective classification model with 97 characteristics instead of the many reclassification models used by prior approaches, and it does not need any additional hardware. These benefits make it a viable and alluring option for organizations to consider.

Many earlier studies focused on passive liveness detection techniques to solve the smart speaker spoofing attack weakness. To differentiate between the speech of a genuine living person and the replayed voice, the detection approach only depends on audio file analysis rather than the use of sensors. However, this detecting system has stringent guidelines for user motions and is significantly influenced by surrounding conditions. Meng et al.~\cite{Meng2022} developed a unique active feature fingerprint that uses the microphone array found in smart speakers to identify the source of audio data gathered. First, they theoretically show that the array fingerprint outperforms previous systems in terms of robust performance regarding environmental changes and human mobility by using the circular architecture of the microphones. To use these fingerprints, they develop characteristics that cooperate with array fingerprints and suggest ARRAYID, a simple passive detection approach. ARRAYID achieves an accuracy of 99.84\%, which is significantly higher than other passive liveness detection techniques. The dataset used for this study had 32,780 audio samples and 14 spoofing devices.

\section{Privacy Issues beyond Technical Attacks}
\label{sec:pri}
IoT device users may not constantly be aware of when and who can access their recordings. Many smart products have built-in microphones, including TVs, doorbells, air conditioners, refrigerators, and others. To prevent consumers from being bugged, a device known as a Protective Jamming Device (PJD) is generally put on top of the speakers of smart voice assistants. By directly pumping interference signals into the smart speaker's microphone, PJD stops eavesdropping. Nevertheless, even with PJD, there is still a significant likelihood of voice eavesdropping because modern voice assistant signal processing procedures cut down on background noise and improve the speech section of the audio. If a hacker gained access to or helped enable recordings made by a smart speaker, they could still stop someone from speaking by successfully canceling the sounds.

Walker et al.~\cite{Walker2021} investigated whether white Gaussian noise (GWN) works well as a PJD interferer. They built a PJD for their experiments and ran attack simulations. To create a store of 1,500 recovered audio recordings, extensive data collecting was carried out in various circumstances. Following data processing, signal/speech quality analysis, speaker and gender detection, time and frequency domain checks, and assessment of metrics such as cross-correlation, SNR, and PESQ, feature extraction (MFCC) was carried out to develop machine learning classifiers. The RandomForest classifier's ability to be utilized to enhance specific voice recognition tasks was determined through experimental findings (beyond the accuracy obtained from random guessing). Another discovery was that 65 dB of source speech allowed for the best accuracy in every circumstance. They hypothesize that improved noise cancellation effectiveness on user's voice closer to 65 dB may be one of the factors contributing to the successful categorization of 65 dB samples. As 70 dB is the loudest point outside the range of typical human speech, noise cancellation may thus be less effective for the user voice at that volume. Due to the increased SPL (power) of the speech signal, which is mistaken for noise, speech-related characteristics can be filtered.

Walker et al.~\cite{Walker2021} attempted to identify songs using the Shazam app. They could attain classification accuracy greater than random guessing for all voice recognition tasks in the studies, such as 46 percent for digit recognition, 51 percent for SI, and 80 percent for gender recognition. Even when PJD secures smart speakers, attackers might still partially utilize recordings made by them to obstruct user speech. Mitev et al.~\cite{Mitev2020LeakyPick} introduced an architecture called LeakyPick to check the smart device communication flow in the cloud. LeakyPick rated 94 percent accuracy while recognizing audio signals from eight voice-assistant-capable devices, and it is cheap to build on a Raspberry Pi. Many techniques stop a device from recording and distributing audio without the user's consent depending on whether LeakyPick serves as a gateway to a home network or passively sniffs (encrypted) Wi-Fi data. An inaudible microphone jamming approach can be used to stop a device from recording private user chats if it can only passively sniff encrypted MAC-layer communication.

LeakyPick was used to find 89 phrases that the Amazon Echo Dot mistakenly identified as wake words, resulting in unexpected audio uploads. However, several of the recognized Alexa wake phrases were only functional when using Google's TTS voice without accurately mimicking human speech. The authors speculate that this could be because TTS-served audio files and real speech behave differently. This experiment illustrates how words were misconstrued as wake words, but it failed to pinpoint the precise phrases that caused incorrect wake words. Mitev's architecture in \cite{Mitev2020LeakyPick} is an affordable approach for the typical user to keep an eye on sound-activated gadgets that unintentionally upload audio segments to the cloud. Further study might investigate the potential for voice assistants and other speech recognition programs to be vulnerable to attacks employing incorrectly identified wake words.

The third party's threats have been revealed in the attack against the ASR system. The third-party market is expanding as a result of the popularity of voice assistants. Young et al.~\cite{Young2022} aimed to protect young users from being tricked and negatively impacted by harmful third-party features. A full dynamic analysis of voice applications with an interactive testing program named SKILLDETECTIVE was conducted to discover speech applications that violate policies on existing VPA systems. SKILLDETECTIVE assesses speech applications against 52 distinct policy requirements in various data formats (text, images, and audio files) to explore voice application behavior automatically and find any policy breaches. VPA platform providers can use SKILLDETECTIVE to proactively identify and stop the release of policy-violating skills at the submission stage. Young et al.~\cite{Young2022} conducted 54,055 tests with Amazon Alexa skills and 5,583 with Google Assistant actions, gathering 518,385 text outputs, about 2,070 unique audio files, and 31,100 unique pictures via voice app interactions. Lastly, 6,079 talents and 175 activities that may contravene at least one policy criterion were found.

Nevertheless, SKILLDETECTIVE has several drawbacks. Due to chatbots' rapid and continual development, the security policy cannot be updated and polished in time, causing their instruments to overlook security infractions. Policy violation detectors are susceptible to adaptive adversary evasion attempts, so malicious developers may modify policy-violating outputs to evade SKILLDETECTIVE's detection. SKILLDETECTIVE ignores some policies (e.g., providing misleading information on skills). SKILLDETECTIVE does not look for abilities that need account connecting to access (for example, skills in the SmartHome category). Simple methods like FNN and bag-of-words were employed to construct chatbots. Future research will replicate this methodology utilizing more sophisticated sentence embeddings like BERT.

Voice assistants are becoming increasingly popular among young people. Le et al.~\cite{Le2021} develop a natural language processing (NLP) system to automatically interact with the voice assistant application and analyze the resulting dialogue to identify risky content for children. The focus was to measure the risk to children of harmful or inappropriate information that may be contained in the voice assistant skills market. 28 kid-targeted applications with questionable material were identified, with a growing collection of 31,966 distinct app behaviors gathered from 3,434 Alexa apps. Their findings imply that kids are susceptible to harmful information despite stricter regulations and greater examination of voice applications made for kids. The user research results revealed that parents were more worried about VPA applications with objectionable material than those that requested personal information. However, many parents were unaware of the dangers of both types of apps. Finally, they discovered a brand-new danger for VPA app users: perplexed vocal commands or utterances shared by several applications, which can cause people to activate or engage with several apps. 4487 obscured utterances were found, and 581 shared both child-directed and non-child-directed apps.

Voiceprints are usually used to authenticate voice assistants. A distinctive manifestation of the human voice created by the spectrum is called a voiceprint. He et al.~\cite{He2022Voiceprint} investigate three key issues to determine if the words affect the distinctiveness of their voiceprints. In particular, it investigates the voiceprints' individuality, the distinctiveness of a voiceprint for a particular word, and the method of creating a wake-up phrase with voiceprint discrimination. Research studies using 2,457 speakers and 14.6 million test samples deconstructed voiceprints into phonemes and discovered relationships between misrecognition rate and phoneme type richness, order, length, and constituents. They devised a phonetic uniqueness computation and gave the output the PROLE Score. They examined 30 wake words from 19 voice assistants available on the market using this new metric to represent the distinctiveness of voices, and they made recommendations for better wake word selection to increase the safety of voice assistants.

The voice assistant's wake-up procedure depends on keyword research. There are two stages to keyword research. The voice assistant is validated and activated by a second cloud model once the on-device model initially identifies keywords in speech samples. The privacy and security concerns that can surface during the procedure are examined in Ahmed et al.~\cite{Ahmed2022}. When the voice assistant was accidentally activated, minute-long audio recordings were transferred to the cloud. Adversarial instances could cause false activations. A remedy named EKOS (Ensemble for KeywOrd Spotting) uses KWS tasks' semantics to protect against unintentional and adversarial activations. During training and inference, EKOS considers the spatial redundancy provided by the auditory environment to reduce distribution drift that might result in unexpected activations.

To deploy a collection of models trained on various harmonics, EKOS uses the physics of speech, namely its redundancy at various harmonics. It provably requires an opponent to change the spectrum for adversarial cases. EKOS maintains natural accuracy while raising the cost of adversarial activation using two complementary strategies to handle the adversarial and inadvertent types of misactivation. Both approaches use various sensors and models to perceive the surroundings. Adversarial activation necessitates a provably cost-increasing guarantee, attained through the physics of audio waves (speech copies of various harmonics) instead of using an ML strategy.

While using a voice assistant application, users are continually monitored by the PVA that communicates with the cloud or other service providers. User control over this process is limited. Given the ubiquitous use of PVA and the solid objection to recording conversations without authorization, more effective recording control must be made accessible. Cheng et al.~\cite{Cheng2019} investigated methods for incorporating more data into acoustic signals for PVA processing. The user applies a marking tool to produce an audio indication when a PVA activity is believed to occur. This identifier will be included in the audio streams that any active PVA records. Tags inform a partner PVA or backend system that the user has not granted their approval for the recording. The recording's time and location can be determined using the tag. Several tagging methods and application situations were compared before creating a prototype PocketSphinx-based tagging gadget. The prototype shows the usefulness of the tagging technique using the well-liked PVA Google Home Mini. This technique may now be used, for instance, to voice complaints to records. Users can turn on the marking device to prevent logging. The tags will be embedded when the PVA is activated, and the backend may delete tag detection data.

Owing to the quick adoption of voice assistants in everyday life, the privacy policy of a voice assistant is a crucial component of keeping users' information private. Although the app store permits third parties to publish their voice apps, developers must supply an acceptable privacy policy. The validity, viability, and degree to which these privacy rules assure user privacy protection are currently unknown. The efficiency of such public privacy policies has not been rigorously measured until Liao et al.~\cite{Liao2020PrivacyPolicies}. They looked at 16,002 Google Assistant actions and 64,720 Amazon Alexa skills. A user study was conducted to learn consumers' thoughts about voice app privacy regulations. Several voice applications suffer from inaccurate privacy, policy URLs, or broken links for the 17,952 skills and 9,955 actions with privacy policies. There are discrepancies in abilities and actions between privacy regulations and descriptions. 6,047 Google activities are needed to contain a privacy statement but do not. Even certified voice applications from Google and Amazon violate their privacy regulations. Their findings were confirmed by the Google Assistant and Amazon Alexa teams.

\section{Challenges and Future Research Directions}
\label{sec:cha}


Because the backdoor attack's performance has been relatively good, future studies should concentrate more on defensive strategies. Identifying the backdoor channels has always been a priority when mitigating backdoor attacks. Most existing defenses against backdoor attacks rely on identifying tainted inputs, which is not always reliable because of well-hidden triggers in sparse data. Retraining is resistant to backdoor attacks, but it is not practical because of its high computational time and indefinite calls to clean training data. In addition, it is presumable that the attacker taints the data in a ``normal" manner by altering the training dataset. If the dataset is poisoned from a distance, it remains unknown whether such an inaudible attack is effective.

Several facets of voice squatting and voice masquerade attacks may benefit future research. These two strikes are new and constantly evolving due to the quickly expanding third-party skill market. The main obstacle to these attacks is the development of an autonomous retrieval system to search for prospective ``susceptible" talents to hijack or fake. The existing approach performs admirably, but the searching step places undue reliance on human effort.

Compared to other attacks, the adversarial attack is one of the most developed attacks in speech. For the adversarial attack, there are still restrictions and difficulties. First, in contrast to the Dolphine Attack, the disturbance introduced would change the audio file and be audible to human hearing. In other words, a human defense might identify the antagonistic samples if he/she listened to every audio clip played. Thus, improving the transcript's plausibility without compromising its correctness remains a future work. It would stop human defenders from being suspicious enough to inspect the audio file. Second, the majority of adversarial attack methods include teaching the model to discover how to swap out or substitute certain target phrases, but it typically fails when encountering untrained or unseen samples. Third, unlike white box settings, the black box setting in adversarial attacks is significantly less investigated.

Most research in adversarial attacks focused on CNN and DNN models. For other machine learning areas, combining transformer ideas into their existing models to improve efficiency is a mainstream trend for end-to-end models. In the last four years, there has been little progress using the advantage of transformer models, except Carlini et al.~\cite{carlini2018audio} in 2018. Even with an impressive attack success rate, Carlini's study was based on white-box knowledge, which did not hugely raise the attack's performance compared to other white-box research. The defense side of the adversarial attack has been greatly developed over the years. Future studies should consider if original sentences can be extracted from adversarial samples after adding noise to lower the adversarial samples' accuracy, for example, Park et al.~\cite{park2021detecting} in 2021. After testing various Gaussian noise distributions, they could retrieve the original transcription findings when the perturbation signal in the opposing samples was minimal. The sound is louder while making huge perturbations of adversarial samples to retrieve the actual transcription from the logit noise. Additional research is required to reconstruct the original transcript using logit noise.

Regarding hidden command attacks, the method proposed in \cite{schonherr2018adversarial} has achieved good attack rates using psychoacoustic concealment. A good attack rate was maintained without affecting human understanding of the original speech samples. However, its main problem is that the injecting sample took a relatively long time compared to other methods. Moreover, it was not tested on black-box experimental environments. The transferability and robustness need to be enhanced. To make the attack more stable and transferable, future research should focus on decreasing the influence of adversarial training mitigation and avoiding transition loss over distance and medium. Another problem in this area is that the defense side was much weaker than the attack, making attacks more dangerous. With the very effective defense method introduced, voice attackers would face a challenge since it calls for them to imitate acoustic fingerprints in both the audio and vibration domains. The playback procedure needed in the detection process might result in front-end mode incursion and some playback delays. Future research may investigate whether and how playback noises and human voices are separated in the vibration domain.

A dolphin attack takes advantage of the microphone's hardware flaw, making it difficult to mitigate without hardware modification. Future research should pay attention to how to change the structure of the microphone and make the change more adaptable and cheaper to adapt. Dolphin attack was introduced in 2017 \cite{Zhang2017Dolphin}. For mitigation, some researchers have raised some ideas of modifying the existing microphone, for example, baseband cancellation and microphone augmentation. However, none of the defenses was adopted by the market. One reason could be that the defense only applies to single-attack mitigation. It is not worthwhile if the dolphin attack was used on a small scale. On the attack side, the attack over the air was mature enough; thus, much research focused on different transmission mediums. With good performance on solid medium, this attack was too specific and could only target specific users, and the attacker himself must have physical access to the attack environment or be close enough to the targeted object.

One of the most significant problems for the third-party market is the policy restriction to protect young users. With increasingly more young users storming into the voice assistant market and the average age of the users being increasingly younger, there are few regulations to distinct the accessibility of the content for adults and kids. The legibility of the skills released to the third-party market was not thoroughly checked due to a lack of restrictions for the service providers. The enforcement of automatic checks for the legibility of every skill before publishing is necessary to safeguard voice assistant applications. Adding a kid's friendly mode and requiring parents' authorization when detecting young users is also essential for voice assistant applications.

In summary, a few aspects should be refined to mitigate technical and non-technical threats in voice assistant applications. This section reviewed challenges and possible future research directions in backdoor attacks, voice squatting and voice masquerade attack, adversarial attack, hidden command attack, and dolphin attacks. The challenges and open issues include attack and defense sides for each attack. This section reviewed the security and privacy problem beyond technical attacks. Policy and supervision of the market are also important for protecting users.

\section{Conclusion}
\label{sec:con}

An in-depth analysis of voice assistant security is provided in this article, with an emphasis on the attacks that may cause a voice assistant to act maliciously and the defenses against such attacks. We start by outlining the overall organization and process of a voice assistant. Then, based on the attacker's intention, threat model, attack strategy, and actual effectiveness, we briefly review several attacks. The six different technological attacks --- backdoor attack, spoofing attack, adversarial attack, hidden command attack, and dolphin attack—are systematized and expanded upon. We categorize current countermeasures into two groups --- detection and prevention—and organize them according to the types of assaults they were intended to avoid and their usefulness in terms of implementation costs, usability, and effectiveness. We also discuss the prospective avenues and current difficulties in future research.

\bibliographystyle{ACM-Reference-Format}
\bibliography{sample-base}


\end{document}